\documentclass[]{emulateapj}
\usepackage{apjfonts}
\usepackage{color}
\newcommand{\hay}{1}
\newcommand{\cfa}{2}
\newcommand{\perimeter}{3}
\newcommand{\waterloo}{4}
\newcommand{\steward}{5}
\newcommand{\mpifr}{6}
\newcommand{\naoj}{7}
\newcommand{\tokyo}{8}
\newcommand{\kasi}{9}
\newcommand{\asiaa}{10}
\newcommand{\bonn}{11}
\newcommand{\asiaahawaii}{12}
\newcommand{\radboud}{13}
\newcommand{\cso}{14}
\newcommand{\mpe}{15}
\newcommand{\berkeley}{16}
\newcommand{\unam}{17}
\newcommand{\astron}{18}
\newcommand{\jcmt}{19}
\newcommand{\guas}{20}
\newcommand{\cita}{21}
\newcommand{\ovro}{22}
\newcommand{\inaoe}{23}
\newcommand{\gent}{24}
\newcommand{\auckland}{25}
\newcommand{\leiden}{26}
\newcommand{\brandeis}{27}
\begin{document}

\title{Persistent Asymmetric Structure of Sagittarius~A* on Event Horizon Scales}
\shorttitle{Persistent Asymmetric Structure of Sgr~A*}
\author{Vincent L.\ Fish\altaffilmark{\hay},
        Michael D.\ Johnson\altaffilmark{\cfa},
        Sheperd S.\ Doeleman\altaffilmark{\hay,\cfa},
        Avery E.\ Broderick\altaffilmark{\perimeter,\waterloo},
        Dimitrios Psaltis\altaffilmark{\steward},
        Ru-Sen Lu\altaffilmark{\hay,\mpifr},
        Kazunori Akiyama\altaffilmark{\naoj,\tokyo,\hay},
        Walter Alef\altaffilmark{\mpifr},
        Juan Carlos Algaba\altaffilmark{\kasi,\asiaa},
        Keiichi Asada\altaffilmark{\asiaa},
        Christopher Beaudoin\altaffilmark{\hay},
        Alessandra Bertarini\altaffilmark{\mpifr,\bonn},
        Lindy Blackburn\altaffilmark{\cfa},
        Ray Blundell\altaffilmark{\cfa},
        Geoffrey C.\ Bower\altaffilmark{\asiaahawaii},
        Christiaan Brinkerink\altaffilmark{\radboud},
        Roger Cappallo\altaffilmark{\hay},
        Andrew A.\ Chael\altaffilmark{\cfa}, 
        Richard Chamberlin\altaffilmark{\cso},
        Chi-Kwan Chan\altaffilmark{\steward},
        Geoffrey B.\ Crew\altaffilmark{\hay},
        Jason Dexter\altaffilmark{\mpe},
        Matt Dexter\altaffilmark{\berkeley},
        Sergio A.\ Dzib\altaffilmark{\mpifr,\unam},
        Heino Falcke\altaffilmark{\radboud,\mpifr,\astron},
        Robert Freund\altaffilmark{\steward},
        Per Friberg\altaffilmark{\jcmt},
        Christopher H.\ Greer\altaffilmark{\steward},
        Mark A.\ Gurwell\altaffilmark{\cfa},
        Paul T.\ P.\ Ho\altaffilmark{\asiaa},
        Mareki Honma\altaffilmark{\naoj,\guas},
        Makoto Inoue\altaffilmark{\asiaa},
        Tim Johannsen\altaffilmark{\cita,\waterloo,\perimeter},
        Junhan Kim\altaffilmark{\steward},
        Thomas P.\ Krichbaum\altaffilmark{\mpifr},
        James Lamb\altaffilmark{\ovro},
        Jonathan Le\'{o}n-Tavares\altaffilmark{\inaoe,\gent},
        Abraham Loeb\altaffilmark{\cfa},
        Laurent Loinard\altaffilmark{\unam,\mpifr},
        David MacMahon\altaffilmark{\berkeley},
        Daniel P.\ Marrone\altaffilmark{\steward},
        James M.\ Moran\altaffilmark{\cfa},
        Monika Mo\'{s}cibrodzka\altaffilmark{\radboud},
        Gisela N.\ Ortiz-Le\'{o}n\altaffilmark{\unam},
        Tomoaki Oyama\altaffilmark{\naoj},
        Feryal \"{O}zel\altaffilmark{\steward},
        Richard L.\ Plambeck\altaffilmark{\berkeley},
        Nicolas Pradel\altaffilmark{\auckland},
        Rurik A.\ Primiani\altaffilmark{\cfa},
        Alan E.\ E.\ Rogers\altaffilmark{\hay},
        Katherine Rosenfeld\altaffilmark{\cfa},
        Helge Rottmann\altaffilmark{\mpifr},
        Alan L.\ Roy\altaffilmark{\mpifr},
        Chester Ruszczyk\altaffilmark{\hay},
        Daniel L.\ Smythe\altaffilmark{\hay},
        Jason SooHoo\altaffilmark{\hay},
        Justin Spilker\altaffilmark{\steward},
        Jordan Stone\altaffilmark{\steward},
        Peter Strittmatter\altaffilmark{\steward},
        Remo P.\ J.\ Tilanus\altaffilmark{\radboud,\leiden},
        Michael Titus\altaffilmark{\hay},
        Laura Vertatschitsch\altaffilmark{\cfa},
        Jan Wagner\altaffilmark{\mpifr,\kasi},
        John F.\ C.\ Wardle\altaffilmark{\brandeis},
        Jonathan Weintroub\altaffilmark{\cfa},
        David Woody\altaffilmark{\ovro},
        Melvyn Wright\altaffilmark{\berkeley},
        Paul Yamaguchi\altaffilmark{\cfa},
        Andr\'{e} Young\altaffilmark{\cfa},
        Ken H.\ Young\altaffilmark{\cfa},
        J.\ Anton Zensus\altaffilmark{\mpifr},
  \&    Lucy M.\ Ziurys\altaffilmark{\steward}
}
\altaffiltext{\hay}{Massachusetts Institute of Technology, Haystack
  Observatory, Route 40, Westford, MA 01886, USA}
\altaffiltext{\cfa}{Harvard-Smithsonian Center for Astrophysics, 60
  Garden Street, Cambridge, MA 02138, USA}
\altaffiltext{\perimeter}{Perimeter Institute for Theoretical Physics,
  31 Caroline Street North, Waterloo, ON N2L 2Y5, Canada}
\altaffiltext{\waterloo}{Department of Physics and Astronomy,
  University of Waterloo, 200 University Avenue West, Waterloo, ON N2L
  3G1, Canada}
\altaffiltext{\steward}{Steward Observatory and Department of
  Astronomy, University of Arizona, 933 North Cherry Ave., Tucson, AZ
  85721-0065, USA}
\altaffiltext{\mpifr}{Max-Planck-Institut f\"{u}r Radioastronomie, Auf
  dem H\"{u}gel 69, D-53121 Bonn, Germany}
\altaffiltext{\naoj}{National Astronomy Observatory of Japan, Osawa
  2-21-1, Mitaka, Tokyo 181-8588, Japan}
\altaffiltext{\tokyo}{Department of Astronomy, Graduate School of
  Science, The University of Tokyo, 7-3-1 Hongo, Bunkyo-ku, Tokyo
  113-0033, Japan}
\altaffiltext{\kasi}{Korea Astronomy and Space Science Institute, 776
  Daedeokdae-ro, Yuseong-gu, Daejeon 305-348, Korea}
\altaffiltext{\asiaa}{Institute of Astronomy and Astrophysics,
  Academia Sinica, P.O.\ Box 23-141, Taipei 10617, Taiwan}
\altaffiltext{\bonn}{Institute of Geodesy and Geoinformation,
  University of Bonn, 53113 Bonn, Germany}
\altaffiltext{\asiaahawaii}{Academia Sinica Institute for Astronomy
  and Astrophysics, 645 N.\ A`oh\={o}k\={u} Place, Hilo, HI 96720,
  USA}
\altaffiltext{\radboud}{Department of Astrophysics/IMAPP, Radboud
  University Nijmegen, PO Box 9010, 6500 GL, Nijmegen, The Netherlands}
\altaffiltext{\cso}{Caltech Submillimeter Observatory, 111 Nowelo
  Street, Hilo, HI 96720, USA}
\altaffiltext{\mpe}{Max Planck Institute for Extraterrestrial Physics,
  Giessenbachstr.\ 1, 85748, Garching, Germany}
\altaffiltext{\berkeley}{University of California Berkeley, Department
  of Astronomy, Radio Astronomy Laboratory, 501 Campbell, Berkeley, CA
  94720-3411, USA}
\altaffiltext{\unam}{Instituto de Radioastronom\'{\i}a y
  Astrof\'{\i}sica, Universidad Nacional Aut\'{o}noma de M\'{e}xico,
  Morelia 58089, Mexico}
\altaffiltext{\astron}{ASTRON, The Netherlands Institute for Radio
  Astronomy, Postbus 2, NL-7990 AA Dwingeloo, The Netherlands}
\altaffiltext{\jcmt}{James Clerk Maxwell Telescope, East Asia
  Observatory, 660 N.\ A`oh\={o}k\={u} Place, Hilo, HI 96720, USA}
\altaffiltext{\guas}{Graduate University for Advanced Studies,
  Mitaka, 2-21-1 Osawa, Mitaka, Tokyo 181-8588, Japan}
\altaffiltext{\cita}{Canadian Institute for Theoretical Astrophysics,
  University of Toronto, 60 St.\ George Street, Toronto, ON M5S 3H8,
  Canada}
\altaffiltext{\ovro}{Owens Valley Radio Observatory, California
  Institute of Technology, 100 Leighton Lane, Big Pine, CA 93513-0968,
  USA}
\altaffiltext{\inaoe}{Instituto Nacional de Astrof\'{\i}sica
  \'{O}ptica y Electr\'{o}nica, Apartado Postal 51 y 216, 72000
  Puebla, Mexico}
\altaffiltext{\gent}{Sterrenkundig Observatorium, Universiteit Gent,
  Krijgslaan 281-S9, B-9000 Gent, Belgium}
\altaffiltext{\auckland}{Auckland University of Technology, 55
  Wellesley Street East, Auckland Central, New Zealand}
\altaffiltext{\leiden}{Leiden Observatory, Leiden University, P.O. Box
  9513, 2300 RA Leiden, The Netherlands}
\altaffiltext{\brandeis}{Department of Physics, Brandeis University,
  Waltham, MA 02454-0911, USA}
\email{vfish@haystack.mit.edu}
\shortauthors{Fish et al.}
\begin{abstract}
The Galactic Center black hole Sagittarius~A* (Sgr~A*) is a prime
observing target for the Event Horizon Telescope (EHT), which can
resolve the 1.3~mm emission from this source on angular scales
comparable to that of the general relativistic shadow.  Previous EHT
observations have used visibility amplitudes to infer the morphology
of the millimeter-wavelength emission.  Potentially much richer source
information is contained in the phases.  We report on 1.3~mm phase
information on Sgr~A* obtained with the EHT on a total of 13 observing
nights over 4 years.  Closure phases, the sum of visibility phases
along a closed triangle of interferometer baselines, are used because
they are robust against phase corruptions introduced by
instrumentation and the rapidly variable atmosphere.  The median
closure phase on a triangle including telescopes in California,
Hawaii, and Arizona is nonzero.  This result conclusively demonstrates
that the millimeter emission is asymmetric on scales of a few
Schwarzschild radii and can be used to break 180\degr\ rotational
ambiguities inherent from amplitude data alone.  The stability of the
sign of the closure phase over most observing nights indicates
persistent asymmetry in the image of Sgr~A* that is not obscured by
refraction due to interstellar electrons along the line of sight.
\end{abstract}

\section{Introduction}
\setcounter{footnote}{0}

\subsection{The Event Horizon Telescope}

Sagittarius~A* (Sgr~A*), located at the Galactic center, marks a dark
mass of just over $4 \times 10^6~M_\odot$
\citep{ghez2008,gillessen2009a,gillessen2009b,chatzopoulos2015}.  At
present there is no credible alternative to a supermassive black hole
\citep{reid2009}.  Its proximity makes it the best studied
astronomical black hole candidate, one for which there is strong
evidence that an event horizon exists \citep{broderick2009b}.  A
variety of observations and theoretical models imply that the
millimeter emission region lies within several Schwarzschild radii of
the black hole ($r_\mathrm{Sch} = 2 G M c^{-2} \approx 10~\mu$as).
Directly resolving the region provides a powerful probe of the
structure and dynamics near the horizon.  General relativity predicts
that Sgr~A* will have a photon ring and associated shadow
approximately $50~\mu$as in diameter
\citep{bardeen1973,falcke2000,takahashi2004}.  Spatially resolved
observations thus hold great promise to assess the nature of the
emission region (e.g., whether the millimeter-wavelength emission
arises from a thick accretion disk or weak jet) as well as to test
general relativity in the strong gravity regime \citep[e.g., via the
  shape and size of the
  shadow;][]{bambi2009,johannsen2010,johannsen2013,broderick2014,psaltis2014,ricarte2015}.

For this purpose the Event Horizon Telescope (EHT) is being assembled.
Comprised of new and existing telescopes at 1.3~mm and 0.87~mm, the
EHT is a global array for very long baseline interferometry (VLBI)
observations of nearby supermassive black holes, including Sgr~A*
\citep{doeleman2009}.  Uniquely among the many telescopes that observe
Sgr~A*, the EHT resolves structures on the scale of a few
Schwarzschild radii in the inner accretion and outflow region.  This
resolution is well matched to the scales of the predicted physical and
astrophysical features.  Previously published EHT observations have
used either the correlated flux density \citep{doeleman2008,fish2011}
or polarization \citep{johnson2015b} on long baselines to infer the
structure of Sgr~A*.  In this work, we focus on a third EHT data
product, closure phases.

\subsection{Closure Phases}

In a radio interferometric array, each baseline produces a complex
observable known as the visibility, which is effectively a Fourier
component of the source image.  The visibility can be decomposed into
two quantities: an amplitude and a phase.  Both parts of the
visibility contain information about the structure of the observed
source.  The amplitude alone can be sufficient to characterize the
approximate size of a source \citep[][and others]{doeleman2008} and
even permit modelling of the source structure
\citep[e.g.,][]{broderick2009,moscibrodzka2009,dexter2010}, but most
of the detailed structural information is contained in the phase
\citep{oppenheim1981}.  For instance, \citet{broderick2011b}
demonstrated that the inclusion of phase data from just a few
telescopes would nail down the spin vector of the black hole in an
accretion flow model of Sgr~A*.

At the high frequencies at which the EHT observes, visibility phases
are easily corrupted by rapidly varying tropospheric delays, primarily
due to water vapor.  A more robust phase observable is the closure
phase, or sum of visibility phases along a closed loop of three
baselines \citep{jennison1958}.  Closure phases are immune to
atmospheric phase fluctuations and to most other phase errors that are
station-based rather than baseline-based in origin, such as phase
variations in the receiver and local oscillator system at each station
\citep{rogers1974}.  Closure phases that are neither zero nor
180\degr\ indicate that the source structure is not point-symmetric at
the resolution of the observing array \citep{monnier2007}.

Nonzero closure phases have been detected on bright quasar sources
with the EHT and used to model the structure of these sources
\citep{lu2012,lu2013,akiyama2015,wagner2015}, but the relative
weakness of Sgr~A* has heretofore only allowed a weak upper limit to
be placed on the absolute value of its closure phase on the
California-Hawaii-Arizona triangle \citep{fish2011}.  In this paper we
report on detections of nonzero closure phases in Sgr~A*, providing
the first direct indication of asymmetric emission near the black
hole.  Multiple measurements of the closure phase of Sgr~A* were
obtained.  We summarize the observing setup and methods of analysis in
Sections~\ref{observations} and \ref{analysis}, describe the results
of the dataset in Section~\ref{results}, examine implications for the
quiescent and variable structure of Sgr~A* in
Section~\ref{discussion}, and comment on future prospects for improved
data in Section~\ref{conclusions}.

\section{Observations} \label{observations}

The EHT obtained detections of Sgr~A* on closed triangles of baselines
among stations in Arizona, California, and Hawaii in 2009, 2011, 2012,
and 2013.  In all cases, two 480-MHz bands, centered at 229.089 and
229.601~GHz (hereafter called low and high bands, respectively), were
observed.  A hydrogen maser was used as the timing and frequency
standard at all sites (but see Section~\ref{2011}).  The two bands
were correlated and post-processed independently.  Digital backends
and phased-array processors channelized each 480~MHz band into 15
channels of 32~MHz each.  Data were recorded on the disk-based
Mark~5B+ and Mark~5C systems \citep{whitney2004b,whitney2010} and then
correlated on the Haystack Mark~4 VLBI correlator \citep{whitney2004}
with a spectral resolution of 1~MHz and an accumulation period of
either 0.5~s or 1~s.  Left-circular polarization (LCP) was always
observed, and right-circular polarization (RCP) was observed in later
experiments as well.  We report only on closure quantities that do not
mix polarizations.

\subsection{Observing Array}

One or more telescopes from each of three sites in Arizona,
California, and Hawaii participated in each set of observations.  The
Arizona Radio Observatory (ARO) Submillimeter Telescope (SMT) on
Mt.\ Graham, Arizona was used in all cases.  At the California and
Hawaii sites, the capabilities of the instruments evolved through the
years, transitioning to recording coherently phased sums of connected
dishes.  Over the years of data analyzed here, the configuration of
VLBI recording at these sites evolved as described below.

The Combined Array for Research in Millimeter-wave Astronomy (CARMA)
in eastern California participated.  Observations always consisted of
two VLBI stations, one of which was a single 10.4-m antenna.  A second
10.4-m antenna participated in 2009 and part of 2011.  From 2011 day
091 onward, the second antenna was replaced by a more sensitive phased
array of up to eight telescopes (including both the 10.4-m and 6.1-m
antennas).  Three observatories on Mauna Kea, Hawaii participated in
observations: the Submillimeter Array (SMA), the James Clerk Maxwell
Telescope (JCMT), and the Caltech Submillimeter Observatory (CSO).
The SMA consisted of a phased array of up to eight telescopes
\citep{weintroub2008,primiani2011}.

Table~\ref{obs-table} summarizes the telescopes participating in each
set of observations along with one-letter station codes, used
hereafter.  Typical fringe spacings are 60~$\mu$as on Hawaii-Arizona
baselines, 70~$\mu$as on Hawaii-California baselines, and 300~$\mu$as
on California-Arizona baselines.

Two stations of the same polarization were used at the CARMA site in
all experiments and on Mauna Kea in 2011.  On arcsecond scales,
extended thermal structures near Sgr~A* contribute to the
millimeter-wavelength interferometer response
\citep[e.g.,][]{kunneriath2012}.  Examination of the correlated flux
density as a function of baseline length indicates that this emission
is resolved out on baselines longer than $\sim 20$~k$\lambda$, or a
projected baseline length of 26~m at $\lambda = 1.3$~mm.  The
intrasite VLBI baselines (CD, DF, EG, JP, and OP) were longer than
$20$~k$\lambda$ except in 2009.

Data quality at 1.3~mm is highly dependent on weather conditions,
which are different from day to day and often variable on any given
day as well.  The sensitivity of the EHT was generally better in later
years due to the inclusion of phased arrays on Mauna Kea and at the
CARMA site.

\begin{deluxetable}{lllll}
\tablecaption{Telescopes Participating in EHT
  Observations\label{obs-table}}
\tablehead{
  \colhead{Station} &
  \colhead{} &
  \colhead{} &
  \colhead{Observing} &
  \colhead{} \\
  \colhead{Letter} &
  \colhead{Telescope} &
  \colhead{Pol.} &
  \colhead{Years (Days)} &
  \colhead{Band(s)}
}
\startdata
C & CARMA (single) & LCP & 2009-2011 (088-090)& both \\
  &                &     & 2011 (091-094)     & low  \\
D & CARMA (single) & LCP & 2009-2011          & both \\
  &                &     & 2012-2013          & low  \\
E & CARMA (single) & RCP & 2013               & low  \\
F & CARMA (phased) & LCP & 2011 (091-094)     & high \\
  &                &     & 2012-2013          & both \\
G & CARMA (phased) & RCP & 2012-2013          & both \\
J & JCMT           & LCP & 2009, 2011 (088)   & both \\
  &                & RCP & 2012-2013          & both \\
O & CSO            & LCP & 2011 (090-094)     & both \\
P & SMA (phased)   & LCP & 2011-2013          & both \\
S & SMT            & LCP & 2009-2013          & both \\
T & SMT            & RCP & 2012-2013          & both
\enddata
\tablecomments{The phased SMA included the CSO on 2011 day 088 and the
JCMT on 2011 days 090-094.  Station F replaced station C in the high
band partway through the 2011 observations.}
\end{deluxetable}

\subsubsection{2009} \label{2009}

Sgr~A* was observed on days 093, 095, 096, and 097, although there
were no detections on the CARMA-Hawaii baselines on day 095.  The
observing array consisted of the SMT, the JCMT, and two individual
CARMA antennas each operating as co-located VLBI sites but using the
same hydrogen maser as a time and frequency standard.  Calibrated
amplitudes from days 095, 096, and 097 have been published in
\citet{fish2011}.

There was significant power aliased into the observing band at the
CARMA stations because the 90\degr\ phase-switching normally used to
separate the sidebands from the double-sideband mixers was disabled
during VLBI scans.  This was not an issue for VLBI baselines between
sites, for which natural fringe rotation was rapid enough to wash out
the contribution from the opposite sideband.  However, the other
sideband was clearly visible in the fringe-rate spectrum on the
intrasite CD baseline, introducing a nonclosing phase error on only
the CD baseline.  Additionally, stations C and D were not separated by
a projected length of 20~k$\lambda$.  As a result of these two
effects, measured closure phases on the CDJ and CDS triangles are
nonzero (see Section~\ref{consistency}) and are therefore excluded
from our analysis.

\begin{figure}
\resizebox{0.49\hsize}{!}{\includegraphics{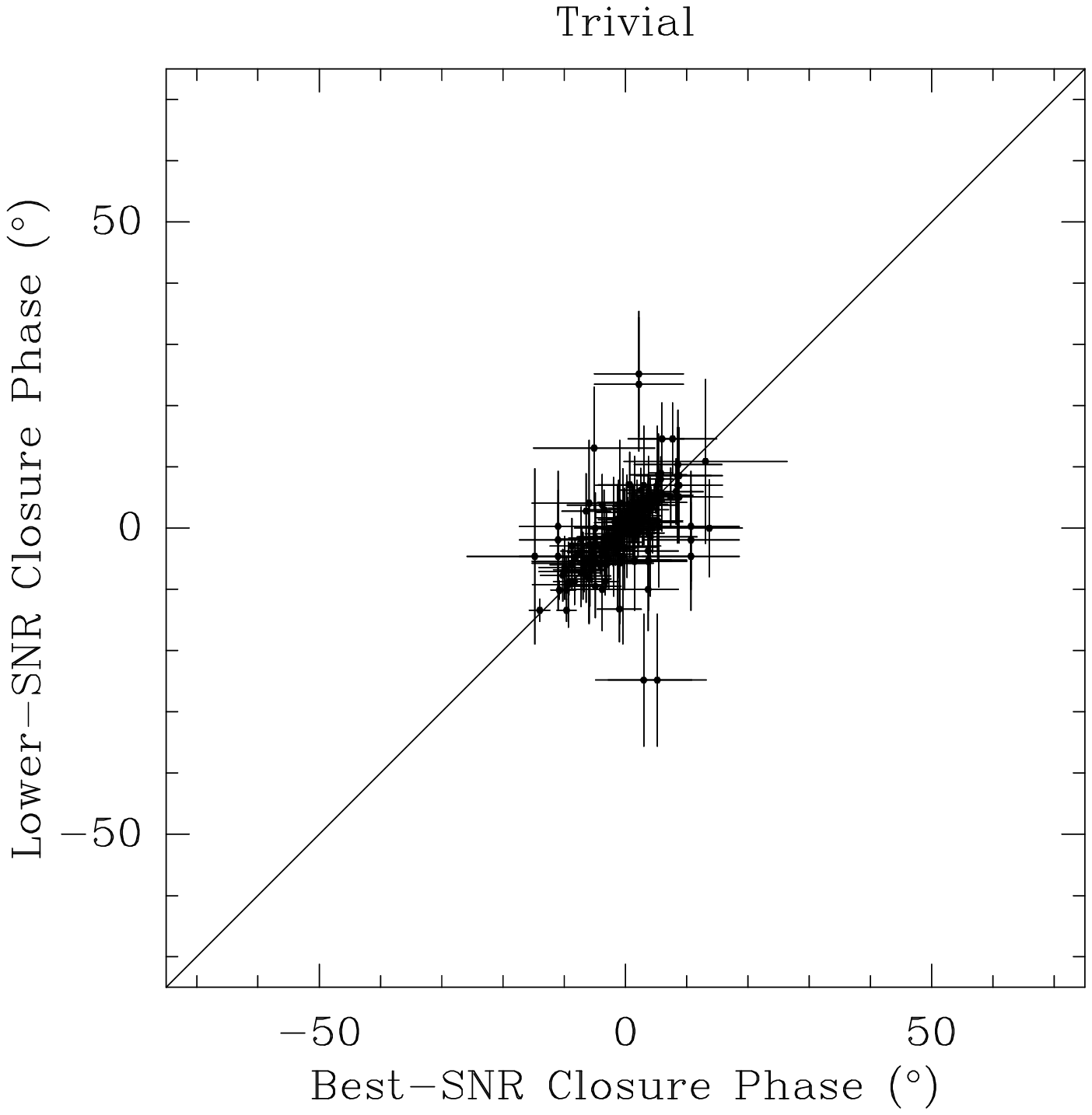}}
\resizebox{0.49\hsize}{!}{\includegraphics{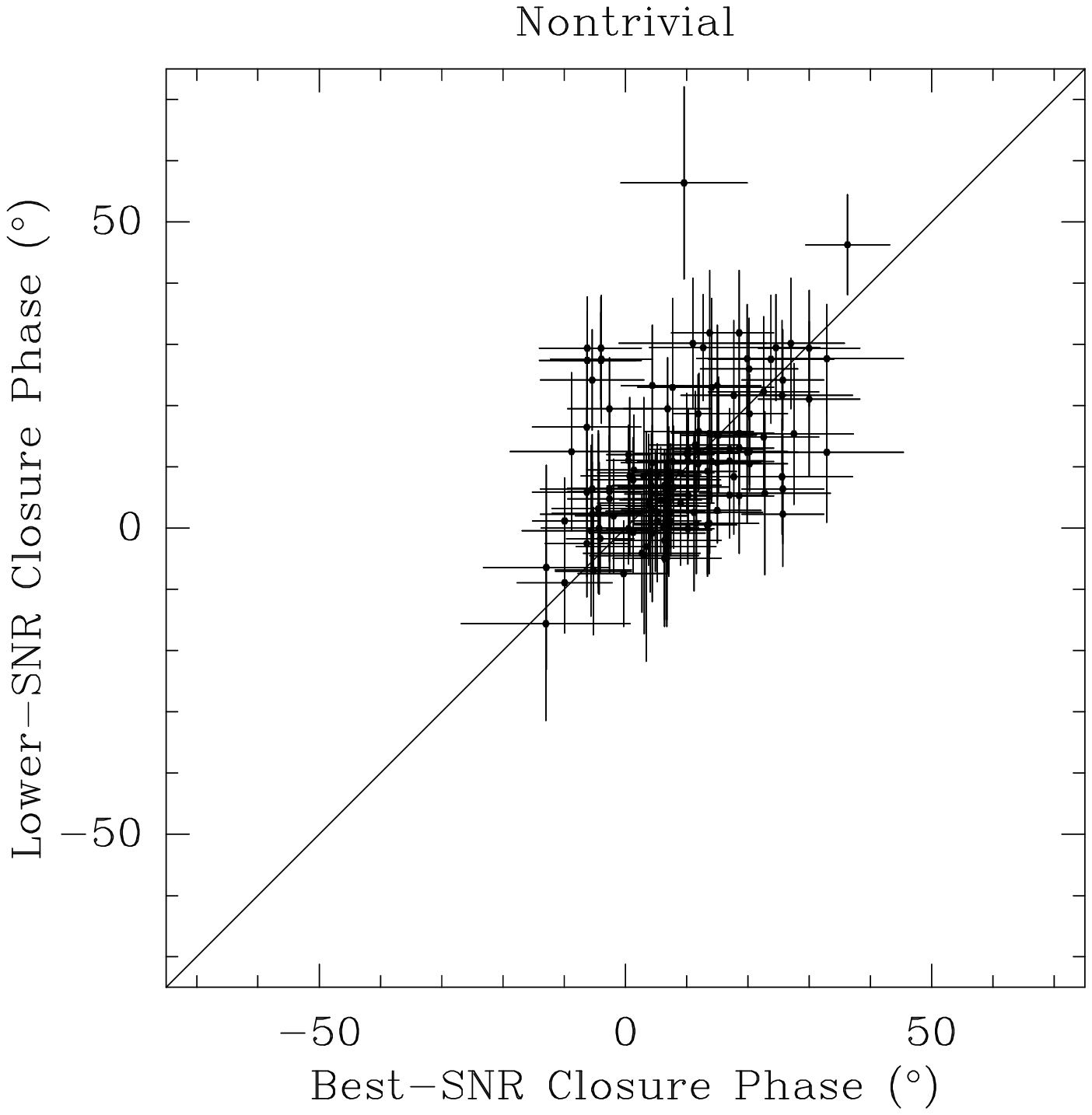}}\\
\resizebox{0.49\hsize}{!}{\includegraphics{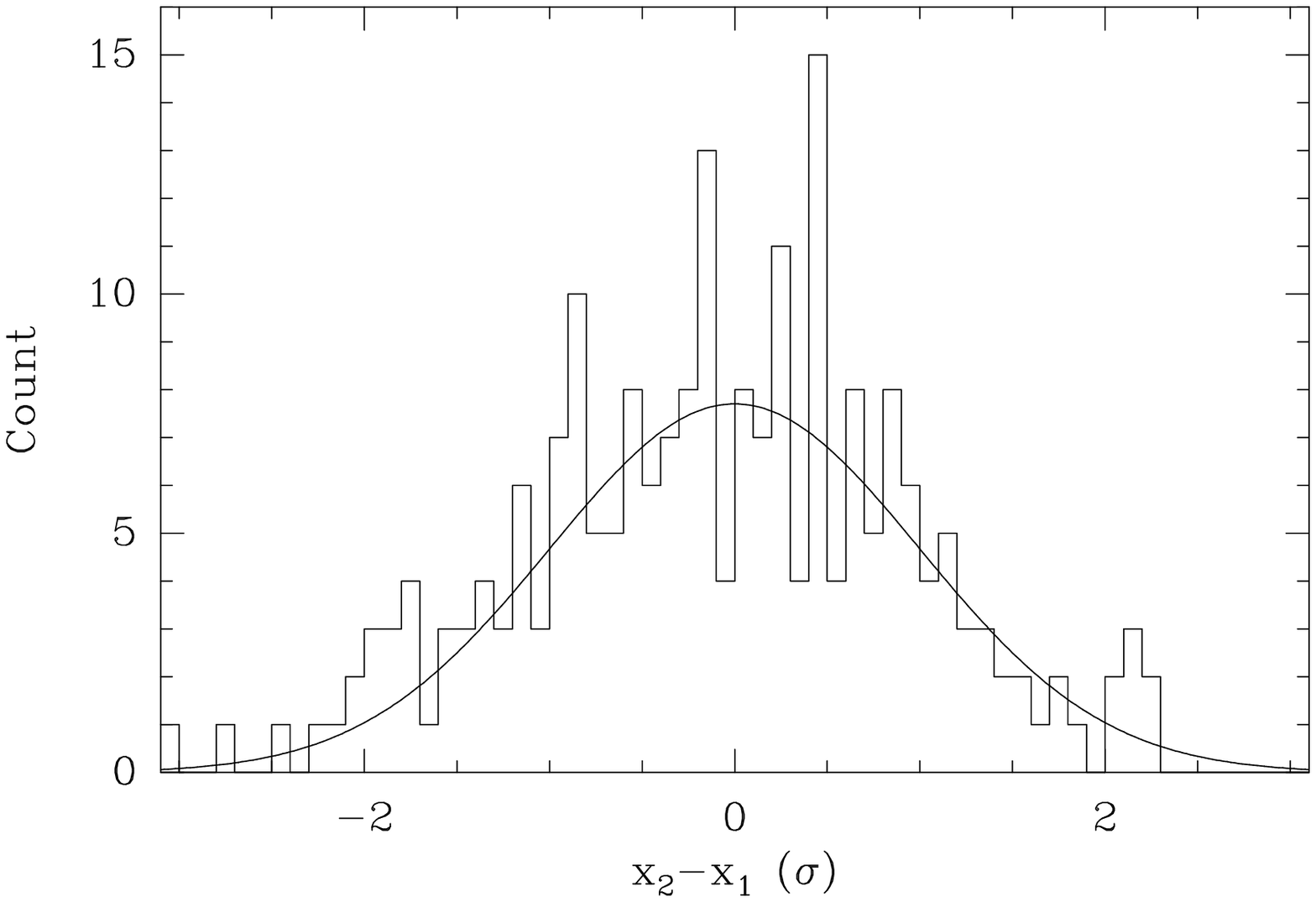}}
\resizebox{0.49\hsize}{!}{\includegraphics{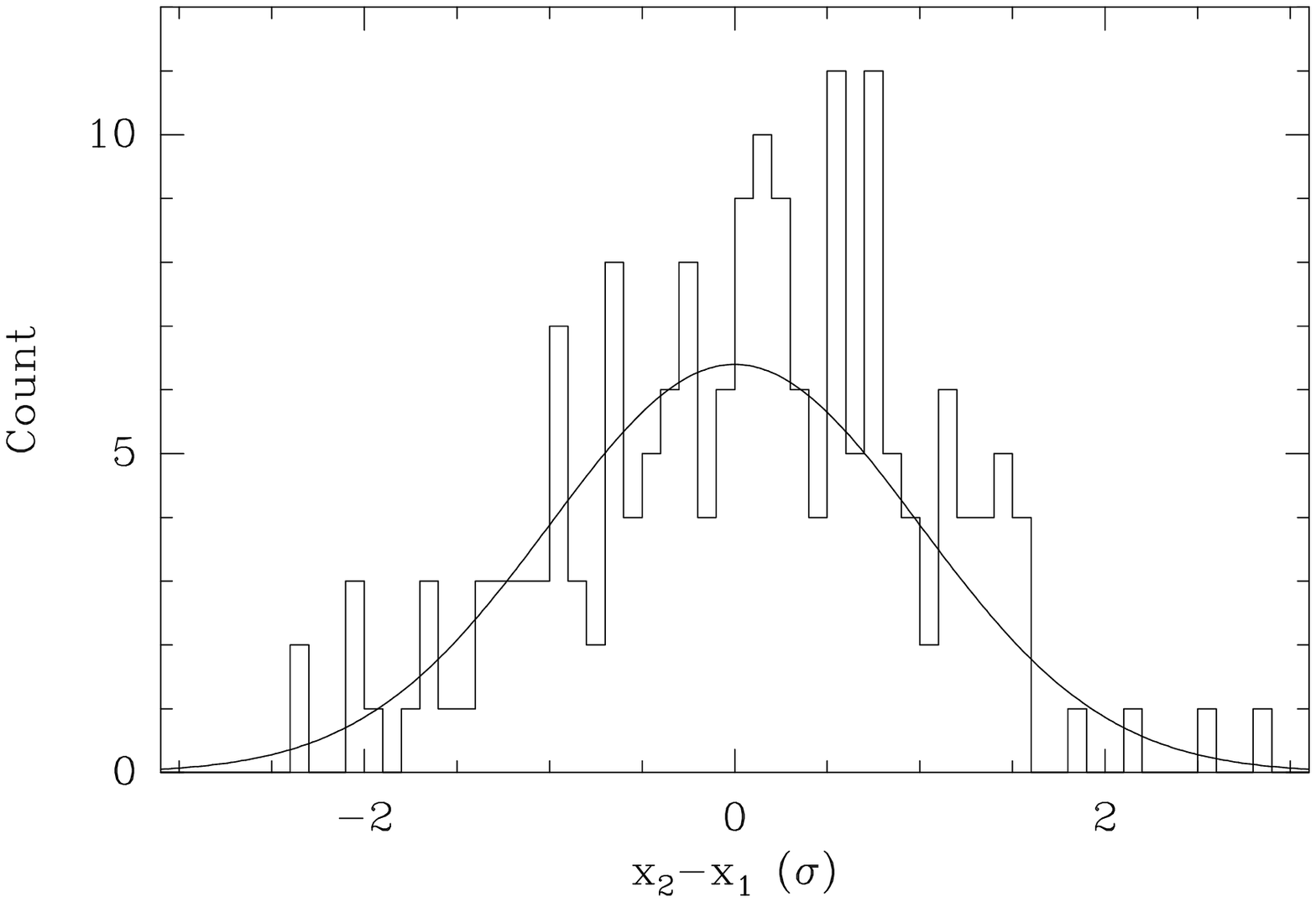}}
\caption{Consistency checks.  Top: Multiple processings of the same
  trivial (left) and nontrivial (right) closure phases are consistent
  to much less than the thermal noise.  Bottom: Pairwise differences
  of similar data points (as defined in Section~\ref{consistency}) are
  consistent with being drawn from a Gaussian random distribution
  characterized by their errorbars, confirming that the error
  estimates are not biased.
\label{fig-consistency}
}
\end{figure}

\subsubsection{2011} \label{2011}

Sgr~A* was observed on days 088, 090, 091, 092, and 094, although day
092 suffered from uncharacteristically high atmospheric turbulence at
the CARMA site.  The observing array consisted of the SMT, two
stations at CARMA, and two stations at Hawaii.  One station consisted
of a single antenna (D).  A second single antenna (C) was used in the
low band on all days and the high band on days 088 and 090.  Station C
was replaced with the phased-array processor (F) in the high band
starting with day 091.  At Hawaii, the JCMT was used as a standalone
antenna on day 088, and the CSO was used on the other days.  The
second station at Hawaii was a phased array that summed signals from
SMA antennas plus either the CSO (day 088) or the JCMT (other days).

While hydrogen masers were used at all sites, the digital backend
sampler clocks, which are the final mix in the signal chain, were
erroneously driven off of the local rubidium clock at CARMA on days
088-092.  An analysis of calibrator sources indicated that this setup
did not affect phase closure.  Further details can be found in
\citet{lu2013}.

\subsubsection{2012} \label{2012}

Sgr~A* was observed on day 075, 080, and 081, although only day 081
provided usable data on all three baselines.  Each site provided
dual-circular polarization observations, with the two polarizations
coming from different telescopes at Mauna Kea.  Disk failures caused
the loss of LCP data from station S in the high band.

\begin{figure}
\resizebox{\hsize}{!}{\includegraphics{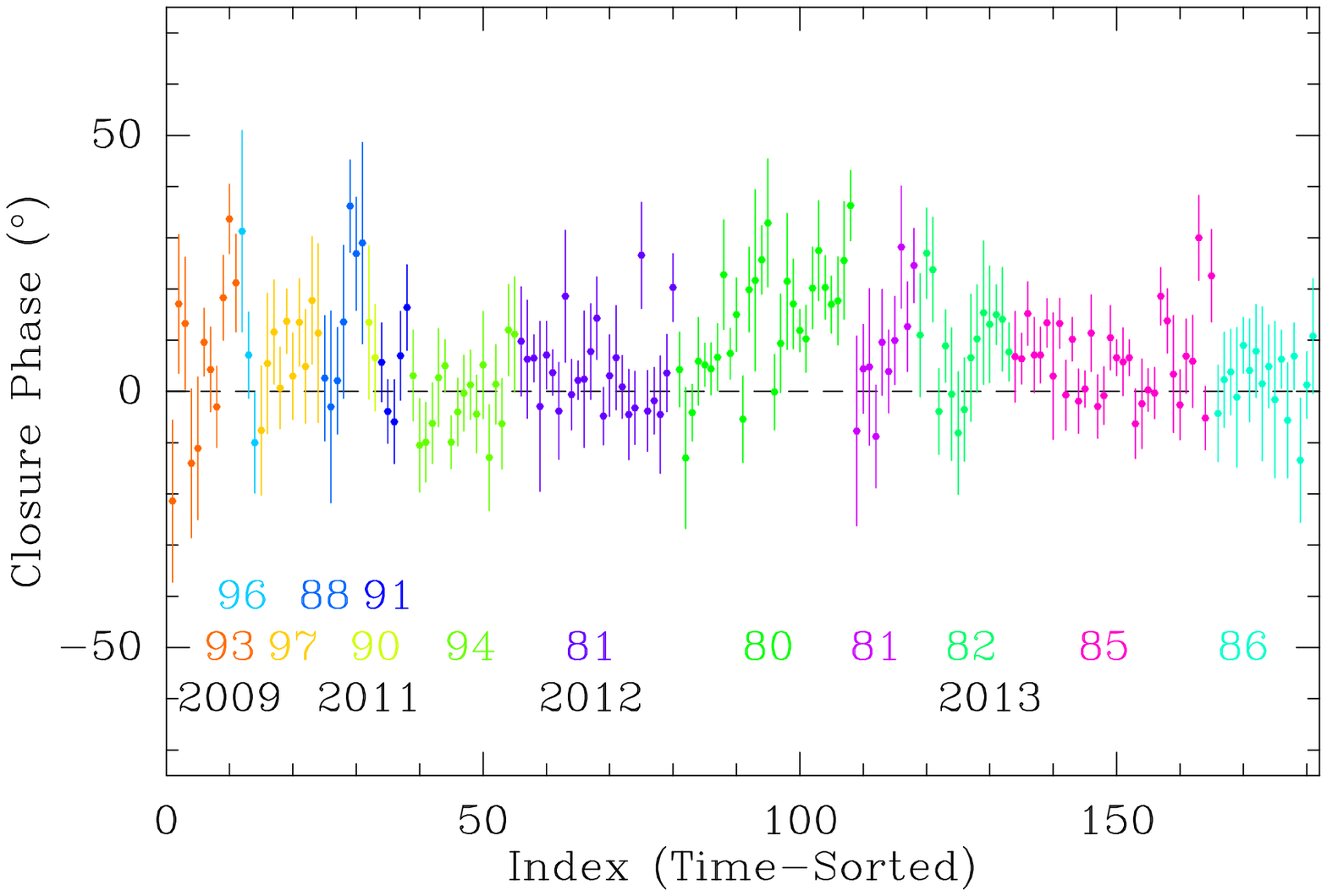}}\\
\resizebox{\hsize}{!}{\includegraphics{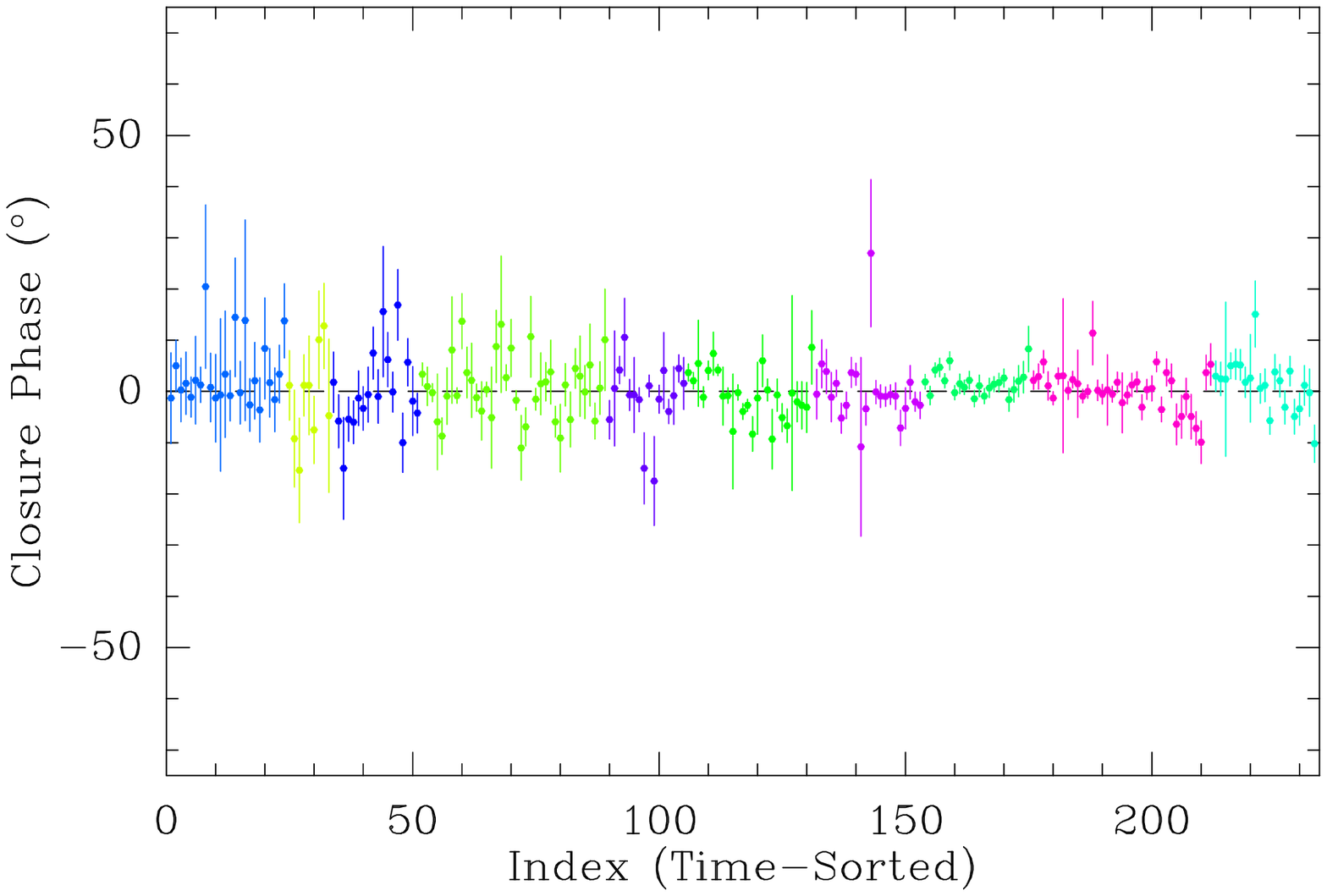}}
\caption{Top: All of the 181 nontrivial California-Hawaii-Arizona
  closure phases measured on Sgr~A*.  Data are presented in time order
  and are color-coded by day and year.  The median nontrivial closure
  phase is $+6.3\degr$.  Bottom: The 233 trivial closure phases for
  Sgr~A*, excluding data from 2009 (Section~\ref{2009}).  The median
  trivial closure phase is consistent with zero, as expected.
\label{fig-data}
}
\end{figure}

\subsubsection{2013} \label{2013}

Sgr~A* was observed on days 080, 081, 082, 085, and 086.  The zenith
opacity at the CARMA site was unusually low, dipping to $0.026$ at one
point, resulting in high sensitivity on the CARMA baselines on some
nights.  The failure of a Mark 5B+ recording system caused the loss of
one polarization in one band at phased CARMA on most nights.
Calibrated visibility amplitudes have been published in
\citet{johnson2015b}.

\subsection{Sign Conventions}

In this work we adopt the sign conventions of \citet{rogers1974} and
\citet{whitney2004}.  The delay on baseline $AB$ is positive if the
signal arrives at station $B$ after station $A$.  A positive delay
produces a positive visibility phase modulo $2\pi$ ambiguities.  The
closure phase on a triangle of three baselines is defined to be the
directed sum of the visibility phases in order: $\phi_{ABC} \equiv
\phi_{AB} + \phi_{BC} + \phi_{CA} = \phi_{AB} + \phi_{BC} -
\phi_{AC}$.

\section{Analysis} \label{analysis}

\subsection{Fringe Search Methods}

Obtaining a closure phase requires detecting the source on a closed
triangle of three baselines.  In practice, source detection is
accomplished by finding a peak in the scan amplitude in a
multidimensional space defined by delays and the delay-rate (residual
to the correlator model values).  The Haystack Observatory
Postprocessing System (HOPS) provides tools to search delay/rate space
and determine the signal-to-noise ratio (S/N) and probability of false
detection associated with the peak.

The rapidly variable troposphere at 1.3~mm introduces large phase
fluctuations on each baseline, introducing challenges for fringe
finding.  When fringes are strong, coherent vector integration along
the entire length of the scan is sufficient to detect the fringe
despite the substantial coherence losses due to the rapidly varying
phase.  Once the delays are well determined, the atmospheric phase
variations are mitigated by segmenting the data at a cadence shorter
than the timescale over which the tropospheric phase changes
appreciably.  Since the tropospheric contribution to the measured
visibility phases close, the visibility phases can be closed on a
per-segment basis and then averaged over the length of the scan to
produce a measurement of the closure phase of the source.  The
bispectral S/N of the resulting averaged closure phase can vary
depending on the choice of segmentation time \citep{rogers1995}.
However, evaluated closure phases at different segmentation times are
self-consistent provided that the segmentation time does not greatly
exceed the coherence timescale.

When fringes are weak, as is often the case on baselines between
Hawaii and the mainland, HOPS supports additional strategies to aid in
fringe detection.  Delay closure can be used to set tight search
windows, aiding in the detection of marginal fringes.  Phase
self-calibration on two strong baselines to a sensitive station can be
used to mitigate atmospheric fluctuations on the third baseline of a
triangle.  Weak fringes can sometimes be detected using incoherent
averaging, in which data are segmented at a cadence comparable to the
coherence time of the atmosphere, and then those segments are
scalar-averaged \citep{rogers1995}.

\subsection{Consistency Checks} \label{consistency}

Since the optimal strategy for fringe detection and closure phase
evaluation varies depending on the sensitivity of each station and
atmospheric conditions, multiple strategies were employed, tailored to
the particular characteristics of each dataset.  When multiple
measurements of the same closure phase were obtained through different
processings, only the data point with the highest bispectral S/N was
retained.  The discarded duplicate points are consistent with the
retained data points (top panels of Fig.~\ref{fig-consistency}),
indicating that the particular methods chosen for fringe detection and
closure phase evaluation do not significantly bias the data.

There are two classes of triangles on which we obtain closure phases.
Triangles that include two VLBI stations from the same site (e.g.,
DFS) should produce closure phases that are trivially zero to within
measurement error.  On the intrasite baseline of these ``trivial''
triangles, the large-scale emission in the Galactic Center (on scales
$\gtrsim 10\arcsec$) is resolved out.  Sgr~A* is then pointlike,
causing the intrinsic source phase to be zero.  The two long baselines
effectively sample the same $(u,v)$ point, adding a source phase on
one baseline and subtracting it on the other.  There is no evidence of
nonzero closure phases on Sgr~A* or other sources in our data on the
trivial triangles.  ``Nontrivial'' closure phases on triangles involve
one CARMA station, one station in Hawaii, and the SMT; these may be
nonzero due to source structure.

As another consistency check, we examined measurements of closure
phases that should be identical to within their errors.  It is
possible that variations in Sgr~A* may cause fluctuations in the
closure phase from scan to scan.  However, during any particular scan,
it is possible to obtain more than one estimate of trivial and
nontrivial closure phases due to duplications among the stations.  The
closure phases obtained in the low and high bands should be identical,
since the fractional frequency difference between the observing bands
is very small.  Closure phases on, e.g., the FPS (LCP) and GJT (RCP)
triangles should be identical, since Sgr~A* exhibits almost no
circular polarization at these frequencies
\citep{munoz2012,johnson2015b}.  Similarly, simultaneous closure
phases on pairs of triangles that share the same sites but with
different stations (e.g., DPS and FPS) should provide measurements of
the same value.  As expected, the pairwise differences of
substantially identical closure phases are consistent with a unit
Gaussian distribution centered on zero when the differences are
normalized by the quadrature sum of the errors of the closure phases
(bottom panels of Fig.~\ref{fig-consistency}).  This also provides
evidence that the closure phase errorbars are correctly estimated.

\section{Results} \label{results}

In total we detect 181 unique nontrivial closure phases for Sgr~A* on
the California-Hawaii-Arizona triangle.  We additionally detect 233
trivial closure phases.  Detected scan-averaged closure phases
  are listed in Table~\ref{data-table}.

The data are shown in Figure~\ref{fig-data}.  There are more data
points in later epochs due to the increased sensitivity provided by
phased SMA and, later, phased CARMA.  Medians of the nontrivial
closure phases are presented in Table~\ref{table-medians} along with
bootstrap estimates of the 95\% confidence interval of the median,
derived from random resampling with replacement.

\begin{deluxetable*}{rrrrrrrrrrrrr}
\tablecaption{Detected Closure Phases\label{data-table}}
\tablehead{
  \colhead{} &
  \colhead{Day of} &
  \colhead{UT Time} &
  \colhead{} &
  \colhead{} &
  \colhead{Closure} &
  \colhead{Bispectral} &
  \colhead{$u_{12}$} &
  \colhead{$v_{12}$} &
  \colhead{$u_{23}$} &
  \colhead{$v_{23}$} &
  \colhead{$u_{31}$} &
  \colhead{$v_{31}$} \\
  \colhead{Year} &
  \colhead{Year} &
  \colhead{(hr)} &
  \colhead{Band\tablenotemark{a}} &
  \colhead{Triangle\tablenotemark{b}} &
  \colhead{Phase ($^\circ$)} &
  \colhead{S/N} &
  \colhead{(M$\lambda$)} &
  \colhead{(M$\lambda$)} &
  \colhead{(M$\lambda$)} &
  \colhead{(M$\lambda$)} &
  \colhead{(M$\lambda$)} &
  \colhead{(M$\lambda$)}
}
\startdata
2009 &  93 & 11.5417 & H & CJS &  -21.4 &    3.78 & -2548.3 & -1691.2 &  3044.2 &  1591.4 &  -495.8 &    99.8 \\
2009 &  93 & 11.9583 & L & CJS &   17.1 &    4.37 & -2658.9 & -1553.0 &  3191.9 &  1425.8 &  -533.0 &   127.2 \\
2009 &  93 & 11.9583 & L & DJS &   13.3 &    4.57 & -2658.9 & -1553.0 &  3191.9 &  1425.8 &  -533.0 &   127.1 \\
2009 &  93 & 12.2917 & L & DJS &  -14.0 &    4.09 & -2724.4 & -1438.7 &  3282.6 &  1288.4 &  -558.2 &   150.3 \\
2009 &  93 & 12.6250 & L & CJS &  -11.1 &    4.25 & -2769.2 & -1322.1 &  3348.2 &  1147.6 &  -579.1 &   174.5 \\
2009 &  93 & 13.1250 & H & CJS &    9.6 &    8.74 & -2796.5 & -1144.7 &  3398.5 &   932.6 &  -602.0 &   212.1 \\
2009 &  93 & 13.4583 & H & CJS &    4.3 &    7.03 & -2788.0 & -1026.1 &  3399.6 &   788.2 &  -611.6 &   237.9 \\
2009 &  93 & 13.4583 & L & CJS &   -3.0 &    7.30 & -2788.0 & -1026.1 &  3399.6 &   788.2 &  -611.6 &   237.9 \\
2009 &  93 & 13.8750 & H & CJS &   18.3 &    6.97 & -2747.3 &  -879.2 &  3364.4 &   608.7 &  -617.0 &   270.5 \\
2009 &  93 & 13.8750 & L & CJS &   33.7 &    8.53 & -2747.3 &  -879.2 &  3364.4 &   608.7 &  -617.0 &   270.5
\enddata
\tablenotetext{a}{High or Low band, as defined in Section~\ref{observations}}
\tablenotetext{b}{Station codes are defined in Table~\ref{obs-table}.}
\tablecomments{Table~\ref{data-table} is published in its entirety in the electronic edition of ApJ.  A portion is shown here for guidance regarding its form and content.}
\end{deluxetable*}

The median closure phase (+6.3\degr) on the California-Hawaii-Arizona
triangle is positive at high statistical significance.  In a larger
run of $10^8$ bootstrap-resampled datasets, every median was positive.
This result is also robust against the exclusion of data from the day
with the largest closure phase (2013 day 080); the resulting dataset
has a median closure phase of 5.0\degr\ with a 95\% lower bound of
3.1\degr.  For comparison, the median \emph{trivial} closure phase is
0.4\degr, consistent with zero (95\% range: $-0.2$\degr\ to
$+1.2$\degr) as expected.  The median nontrivial closure phase of
+6.3\degr\ is too large to be attributable to instrumental effects
(Appendix) and is nearly identical to the +6.7\degr\ measured in an
independent analysis of 2013 data by R.-S.\ Lu et al.\ (in
preparation) using both the Mark~4 and DiFX correlators.

\begin{deluxetable}{ccrrrr}
\tablewidth{0.99\hsize}
\tablecaption{Median Closure Phases of Sgr A* on the California-Hawaii-Arizona Triangle\label{table-medians}}
\tablehead{
  \colhead{} &
  \colhead{} &
  \colhead{} &
  \colhead{} &
  \multicolumn{2}{c}{95\% Range\tablenotemark{a}} \\
  \colhead{Year} &
  \colhead{Day(s)} &
  \colhead{$N$} &
  \colhead{Median} &
  \colhead{Low} &
  \colhead{High}
}
\startdata
2009 & 093 &  11 &    9.6 & $-$11.1 & 17.7 \\
2009 & 096 &   3 &    7.1 & \nodata &\nodata \\
2009 & 097 &  10 &    8.4 &     0.7 & 13.5 \\
2009 & All &  24 &    8.4 &     0.7 & 13.5 \\
2011 & 088 &   7 &   13.6 &  $-$0.4 & 29.0 \\
2011 & 090 &   2 &   10.0 & \nodata &\nodata \\
2011 & 091 &   5 &    5.7 &  $-$5.9 & 11.7 \\
2011 & 094 &  17 & $-$0.3 &  $-$7.2 &  2.7 \\
2011 & All &  31 &    2.6 &  $-$3.5 &  5.7 \\
2012 & 081 &  25 &    3.1 &  $-$1.8 &  6.5 \\
2013 & 080 &  28 &   16.0 &     6.7 & 20.2 \\
2013 & 081 &  10 &    7.2 &  $-$7.7 & 12.7 \\
2013 & 082 &  15 &   10.3 &  $-$0.5 & 14.1 \\
2013 & 085 &  32 &    6.5 &     0.5 &  7.1 \\
2013 & 086 &  16 &    3.0 &  $-$1.6 &  6.3 \\
2013 & All & 101 &    6.9 &     5.6 &  9.4 \\
All  & All & 181 &    6.3 &     4.3 &  7.0
\enddata
\tablecomments{All closure phases are measured in degrees.}
\tablenotetext{a}{The 95\% confidence interval of the median is
  estimated from bootstrap analyses using $10^7$ resampled datasets.
}
\end{deluxetable}

Care must be taken not to overinterpret differences between subsamples
of the dataset.  When individual days of data are processed in
multiple ways, the median closure phase can differ by a few degrees.
This is particularly true for data taken in 2009 and 2011 (before the
sensitivity of the observing array was increased) or for subsamples
consisting of only a few measurements, where the inclusion or
exclusion of one or two marginal detections can have a greater impact
on the derived median value.  For instance, the median closure phase
on 2011 day 094 increases from $-0.3$\degr\ to $+3.5$\degr\ when only
self-calibrated data points are considered.  The dataset taken as a
whole is large enough to be robust against the details of the
processing to within a few tenths of a degree.

\begin{figure}
\resizebox{\hsize}{!}{\includegraphics{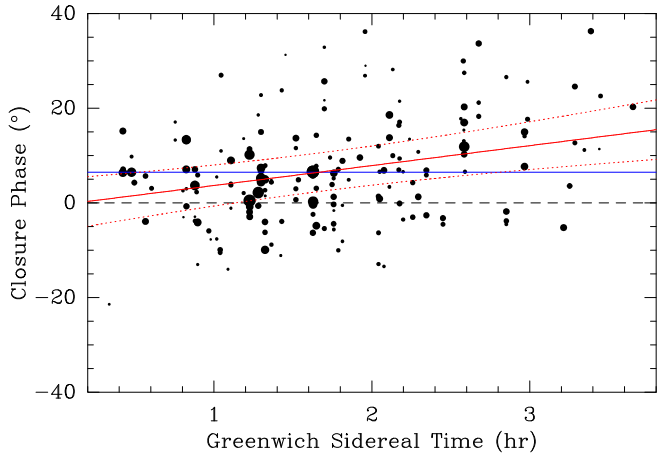}}
\caption{Measured closure phases (dot diameter proportional to S/N)
  plotted against GST.  The solid red line shows the best-fit line,
  with the dashed red lines showing the $\pm3\,\sigma$ range.  The
  solid blue line shows the best-fit line with zero slope.  Despite
  the large scatter, the data suggest that the
  California-Hawaii-Arizona closure phase may be increasing with GST.
\label{cphase-vs-gst}
}
\end{figure}

Since each baseline samples different $(u,v)$ points at different
times due to Earth rotation, it would be possible for the closure
phase to vary as a function of time even if Sgr~A* did not exhibit
variability.  Indeed, there is a trend for the measured
California-Hawaii-Arizona closure phase to be larger later in the
observing track (Fig.~\ref{cphase-vs-gst}).  The $\chi^2$ per degree
of freedom for the best-fit line is 1.43, indicating that there is
additional variability and/or that the increase with time is not
linear.  This increase with GST is significant at a level of
$>\,4\sigma$ using a Kendall tau test.  The expected functional
form as a function of GST is model-dependent, although the general
trend for the California-Hawaii-Arizona closure phase to increase with
time over an observing night provides an important constraint for
physically motivated models of the 1.3~mm emission region in Sgr~A*.

\section{Discussion} \label{discussion}

Our data clearly demonstrate that the closure phase on the
California-Hawaii-Arizona triangle is nonzero, with a trend for the
magnitude of the closure phase to increase over the course of a night.
The sign and approximate value of the closure phase are consistent
among multiple observing epochs over four years.  In this section we
consider the implications of these results.

\subsection{Implications of Nonzero Closure Phase}

The detection of nonzero closure phase is an unambiguous indication
that these EHT data are resolving structure in the image of Sgr~A*.
Two robust conclusions are that the morphology of the emission from
Sgr~A* at 1.3~mm cannot exhibit point symmetry and that the millimeter
emission is asymmetric on scales of a few Schwarzschild radii.

The sign of the closure phase resolves 180\degr\ rotational
ambiguities in models.  As an example, the best-fit parameters for the
\citet{broderick2011} model find that the rotation axis of the
accretion disk points toward either $-$52\degr\ or
$+$128\degr\ ($2\,\sigma$ error $+33\degr/-24\degr$) east of north.
This pair of directions was obtained from visibility amplitude
information alone, which cannot discriminate between the two
directions because the Fourier transforms of an image and the same
image rotated 180\degr\ are identical modulo a sign flip in phase.
Therefore, visibility phase or closure phase information is required
to break the 180\degr\ degeneracy.  The $+$128\degr\ direction is
consistent in sign with our measured closure phases.  Adding the new
closure phase data is likely to result in better estimates of model
parameters and to provide stronger constraints on models, including
those that allow for deviations from general relativity
\citep[e.g.,][]{broderick2014}.

\subsubsection{Accretion Models}

The detection of nonzero but small closure phases in the image of
Sgr~A* demonstrates the power of imaging observations in placing
strong constraints on its accretion flow geometry.  In particular, the
data favor emission morphologies that are connected rather than those
composed of disjoint regions at horizon scales.  At this point, it is
instructive to look at the images generated in different general
relativistic magnetohydrodynamic (GRMHD) simulations in order to
explore in more detail how our observations can be used to constrain
various configurations that are physically plausible and are
consistent with all other currently available data.

The parameters of GRMHD simulations are typically calibrated in order
to reproduce the broadband spectrum of Sgr~A* as well as the overall
size of its emitting region at 1.3~mm. Even with these constraints
imposed, however, the images they generate can be quite different from
each other, depending on the prescription for the plasma
thermodynamics that was employed as well as on the initial magnetic
field configuration and tilt of the torus that was used to feed the
black hole.

In a set of simulations referred to by \citet{narayan2012} as Standard
and Normal Evolution (SANE), the magnetic flux remains modest
\citep[e.g.,][]{devilliers2003,gammie2003}.  If the electron
temperature is assumed to be at a constant ratio with the ion
temperature everywhere in the flow, the generated images typically
show continuous crescent-like brightness distributions
\citep{moscibrodzka2009,moscibrodzka2012,dexter2009,dexter2010,chan2015}.
On the other hand, if the electrons in the jet are allowed to be
heated much more strongly than within the disk, the images are
characterized by bright regions from the inner walls of the jets,
dissected by the cooler accretion disks
\citep{moscibrodzka2013,moscibrodzka2014,chan2015}.  The relative
brightness of the two regions (and their exact shape) depends on the
inclination of the observer.  Images with even more disjoint bright
regions arise in Magnetically Arrested Disk simulations (MAD in the
terminology of \citealt{narayan2012}; see also \citealt{mckinney2009}
and \citealt{dexter2012} for similar magnetically dominated
simulations and their applications to EHT observations of M87) and are
dominated by emission from the footpoints of the jets
\citep{chan2015}.  Finally, images with separated bright regions arise
naturally in GRMHD simulations in which the accreting material is fed
to the black hole from a plane that has a tilt with respect to the
black-hole spin \citep{dexter2013} because of the presence of standing
shocks in these flows.

Comparing the detailed predictions of these simulations to our data is
beyond the scope of the current paper. However, motivated by the
rather general properties of the disjoint geometries in the images
exhibited by some of the GRMHD simulations, we discuss below how the
observations reported here can be used to constrain such
configurations.

\subsubsection{Constraints on Disjoint Bright Regions}

The gross characteristics of the jet-dominated images with disjoint
bright regions described above
\citep[see][]{moscibrodzka2013,moscibrodzka2014,chan2015} can be
captured by a simple geometric model composed of two separated regions
that are symmetric and identical except for a difference in the
brightness of each component.  The closure phases of such a
configuration will be equivalent to an even more reduced model in
which the two regions are replaced with point sources.  In this
reduced model, the visibilities are analytically calculable as $V(u,v)
= 1 + re^{-2 \pi i (ux+vy)}$, where $r$ represents the amplitude ratio
of the two components, and $x$ and $y$ refer to their separation east
and north, respectively\footnote{Since closure phase is
  translation-invariant, we place one component at the origin for
  convenience.}.

\begin{figure}
\resizebox{\hsize}{!}{\includegraphics{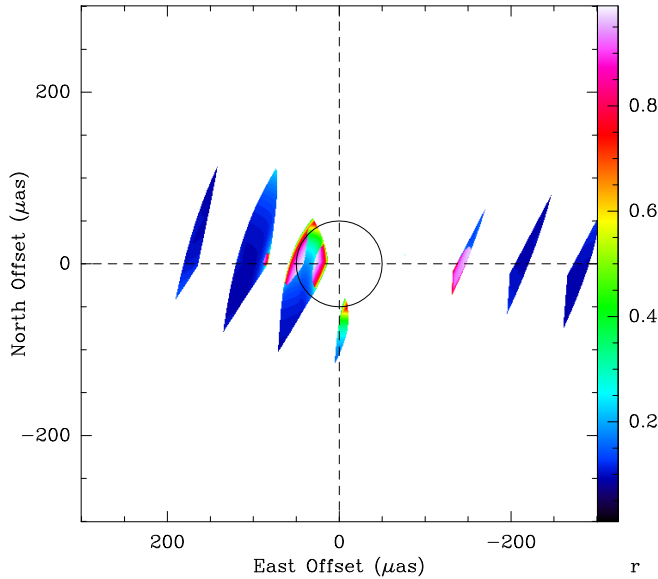}}
\caption{Offsets between two point sources that would produce closure
  phases between +0.9\degr\ and +14.9\degr\ at all triangles of
  $(u,v)$ coordinates sampled by our data.  For a unit point source
  centered at the origin, the colored regions indicate the allowed
  offset of a second point source, with color indicating the maximum
  value of $r$ within the range $0 < r < 1$.  The circle shows an
  offset of 50~$\mu$as, approximately the diameter of the predicted
  shadow in Sgr~A*.  For two components separated by the shadow
  diameter, a roughly east-west alignment with the brighter component
  to the west is required to be consistent with our data.
\label{two-point-sources}
}
\end{figure}

Figure~\ref{two-point-sources} shows the separations between two point
sources that are consistent with our data.  These separations would
produce California-Hawaii-Arizona closure phases that are between
$+0.9$\degr\ and $+14.9$\degr, the range of values implied by the
best-fit line in Figure~\ref{cphase-vs-gst}, at all GST times for
which we have measured closure phases.  Both $r = 0$ (one point
source) and $r = 1$ (two equal point sources) are inconsistent with
our data, since each would produce closure phases that are identically
zero or $180$\degr.  For a separation comparable to the shadow
diameter, the closure phase data imply that a disconnected
two-component model would be oriented roughly east-west with the
brighter component located to the west
(Figure~\ref{two-point-sources}).  If the emission from Sgr~A* is
coming from the footpoints of a MAD-type jet, this is inconsistent
with the orientations of the \citet{li2013} jet or \citet{bartko2009}
clockwise disk (see extended discussion in \citealt{psaltis2015}) but
aligned with the preferred axis of the intrinsic emission at 7~mm
\citep{bower2014}.  Closure phase measurements at 86 GHz already rule
out asymmetric jet structures on larger angular scales from a few
hundred microarcseconds to a few milliarcseconds \citep{park2015}.

\subsection{Consistency of Closure Phases}

\subsubsection{Alignment of the Accretion Disk and Black Hole Spin Axes}

The closure phase on the California-Hawaii-Arizona triangle is
consistent in sign and magnitude, to within measurement error, from
day to day.  These observations span a 4-year timescale that is much
longer than the orbital period at the innermost stable circular orbit,
which ranges from a few minutes to about half an hour depending on the
spin of the black hole.  It is also larger than the Lense-Thirring
precession timescale for a tilted accretion disk unless the effective
outer accretion flow radius is very large or the black hole spin is
very small \citep{fragile2007,dexter2013}.

Misalignment of the spin axes of the accretion disk and black hole
could produce two different observational consequences.  First, it is
possible that the inner disk could have a stable but slowly precessing
structure.  Examination of each epoch of the data in the context of
disk \citep[e.g.,][]{broderick2009,broderick2011} or geometric
crescent models \citep{kamruddin2013} may be able to place limits on
the amount of precession.

Second, misalignment could result in multiple bright regions in the
accretion flow due to standing shocks.  GRMHD simulations of the
accretion flow find that the separation between these regions is
comparable to the diameter of the photon ring, although the emission
pattern can be quite complicated in general \citep{dexter2013}.  Since
the standing shocks can travel faster than the Lense-Thirring
precession speed, the accretion flow would be expected to have
substantially different structure in different years and very likely
on different days within each year.  The predicted closure phases over
the course of a day on the California-Hawaii-Arizona triangle for the
515h model of \citet{dexter2013} mimic the rough range of closure
phases observed, including a general trend of increasing value with
GST.  Further study is required to determine whether such a model
predicts excess variability in closure phases and long-baseline
amplitudes beyond what is observed.  The increased sensitivity of the
EHT in upcoming years will be helpful for examining variability on
intraday timescales.

\subsubsection{Connections with the Accretion Rate}\label{accretionrate}

The discovery of the G2 gas cloud on an orbit with a close approach to
Sgr~A* \citep{gillessen2012} sparked interest in the possibility that
the accretion rate of Sgr~A* would increase due to the introduction of
additional material into the accretion flow.  Most of the material in
G2 did not pass through pericenter until after the final epoch of
observations reported herein \citep{gillessen2013}, and mounting
evidence suggests that G2 contains a star and is therefore not a pure
gas cloud \citep{eckart2014,witzel2014,valencias2015}.  In any case,
the infall timescale is on the order of years \citep{burkert2012}, so
it would not be expected for there to be observational signatures of
the G2 event in these data.  However, there is evidence that G2 is a
knot in a larger gas streamer that also includes the G1 gas cloud,
which reached pericenter in the middle of 2001 \citep{pfuhl2015}.  If
so, the accretion flow of Sgr~A* could be supplemented with material
from G1 or other gas in the streamer, with the caveat that some of the
material deposited in the outer accretion flow may be carried away by
outflows rather than making it to the inner region traced by the
1.3~mm emission \citep{wang2013}.

The consistency of closure phases across multiple epochs from 2009
through 2013 provides evidence against large changes in the accretion
rate over this period, consistent with the results of radio and
millimeter-wavelength monitoring during the G2 encounter
\citep{bower2015}.  The GRMHD simulations of \citet{moscibrodzka2012}
are instructive.  When the accretion rate is decreased, the effective
size of the accretion region decreases, with the result that the
Hawaii-Arizona and Hawaii-California baselines do not adequately
resolve the 1.3~mm emission region, causing the predicted closure
phases to drop very close to zero.  As the accretion rate is
increased, the effective size of the emission region becomes larger,
producing larger closure phases as the fringe spacing of the
Hawaii-Arizona and Hawaii-California baselines becomes better matched
to the asymmetric emission region.  These larger closure phases
persist even at the largest accretion rates modelled by
\citet{moscibrodzka2012}, where the shadow of the emission region is
obscured by the high optical depth of the accretion flow.  This
behavior is also seen when the accretion rate of the
\citet{broderick2011} radiatively inefficient accretion flow models is
varied.  The average California-Hawaii-Arizona closure phase may
therefore provide information complementary to visibility amplitudes
in determining the overall accretion rate of the Sgr~A* system.

\subsubsection{Limits on Refractive Phase Noise}

Scattering in the tenuous plasma of the interstellar medium affects
the image of Sgr~A* at radio wavelengths.  The biggest effect of this
scattering is to blur the image of Sgr~A*, causing its apparent size
to vary approximately as the square of the observing wavelength
\citep[][and many others]{davies1976,lo1981,doeleman2001,bower2006}.
A secondary effect of scattering is to introduce variable substructure
within the scattered image \citep{gwinn2014,johnson2015}.  Both of
these effects can modify VLBI observables.

The formalism of \citet{narayan1989} and \citet{goodman1989}
distinguishes between three different regimes of scattering.  In the
snapshot regime, diffractive scattering from small-scale
inhomogeneities dominates.  In the average regime, diffractive
scattering is quenched, but refractive scintillation from large-scale
inhomogeneities persists.  In the ensemble-average regime, both
diffractive and refractive scintillation are suppressed, and the
scattering produces a deterministic blurring of the image.  Due to the
intrinsic size of Sgr~A* as well as the integration time and bandwidth
used in VLBI observations, the average regime is applicable to EHT
observations of Sgr~A* over the course of a single night.  The
ensemble of many nights of observations will tend statistically toward
the ensemble-average regime, in which the observed visibilities are
the intrinsic visibilities downweighted by a real Gaussian whose width
in baseline space is inversely related to the size of the scattering
ellipse.  In the ensemble-average regime, the effects of
scattering---blurring of the image---are invertible and do not affect
closure phases \citep{fish2014}.

However, in the average regime, image distortions and refractive
substructure can introduce nonclosing phases.  The magnitude of these
effects may be up to 50~mJy in the visibility domain, with peak effect
at baselines near the length at which the ensemble-average visibility
of a point source falls to $1/\sqrt{e}$ \citep{johnson2015}.  This
could be as large as a 10\% effect on top of source visibilities of
approximately 500~mJy on the baselines between Hawaii and the mainland
US \citep{fish2011}, corresponding to a phase noise of $\sim 6^\circ$.
However, this refractive phase noise will be partially correlated when
the antennas are located within a few thousand kilometers of each
other, so the net effect on closure phases on the
California-Hawaii-Arizona triangle will be smaller.

Because refractive phase noise will fluctuate randomly about zero on
timescales of about 1 day \citep{fish2014}, it cannot account for the
nonzero closure phases that we measure, which show a strong tendency
to be positive.  Nor can refractive phase noise account for the
dependence of closure phase on GST.  Thus, our measurements are a
secure indication of intrinsic source asymmetry and do not merely
reflect scattering-induced asymmetrical substructure.  However,
refractive variations may contribute to smaller interday variations in
closure phase.  Further observations at a range of wavelengths will be
required to characterize the properties of the turbulent scattering
screen and better understand its contribution to the apparent
variability of Sgr~A*.

\section{Conclusions and Future Prospects} \label{conclusions}

We have obtained 181 measurements of the closure phase at 1.3~mm on
the California-Hawaii-Arizona triangle from 2009 to 2013.  The median
closure phase is nonzero at high statistical significance.  This
provides the first direct evidence that the structure of the 1.3~mm
emission region is asymmetric on spatial scales comparable to the
diameter of the shadow around the black hole that is predicted by
general relativity.  If the 1.3~mm emission arises from a MAD-like jet
whose emission is concentrated in two disjoint bright regions
separated by the shadow diameter, our data place a very strong
constraint on its orientation.  The data also provide important
constraints for parameters of other outflow and accretion models of
Sgr~A*.

The constancy of the sign of the closure phase argues for the
persistence of asymmetric quiescent structure in Sgr~A* likely coupled
with some structural variability, consistent with simulations of a
dynamic, spin-aligned accretion disk.  While it is not currently
possible to entirely disentangle the effects of variability in the
structure of the emitting material around Sgr~A* from apparent
substructure introduced by variations in the scattering screen, the
California-Hawaii-Arizona closure phases indicate that refractive
phase noise is not dominant on baselines between Hawaii and the
western US.  These results are encouraging for producing an image of
the quiescent emission by averaging several nights of data to mitigate
intrinsic source variability and refractive substructure
\citep{lu2015}.

Closure phase measurements will soon provide much stronger constraints
on source structure.  In the near term, EHT observations in 2015 and
beyond will incorporate additional observatories, including the Large
Millimeter Telescope in Mexico, providing closure phase data on new
triangles with higher angular resolution.  Increasing data rates,
starting with dual-polarization 2~GHz observations in 2015, will
provide increased sensitivity that will result in larger detection
rates and smaller random errors on each closure phase measurement,
allowing interday and intraday variability to be tracked more
accurately.  Completion of the 1.3~mm VLBI array---including phased
ALMA \citep{fish2013}, the South Pole Telescope, the IRAM 30-m
telescope at Pico Veleta, and the Northern Extended Millimeter Array
at Plateau de Bure---will produce very sensitive data with good
baseline coverage, culminating in the ability to reconstruct
model-independent images of Sgr~A*, M87, and other sources
\citep{lu2014}.

\acknowledgments

The EHT is supported by multiple grants from the National Science
Foundation (NSF) and a grant from the Gordon \& Betty Moore Foundation
(GBMF-3561) to S.~S.~D.  The SMA is a joint project between the
Smithsonian Astrophysical Observatory and the Academia Sinica
Institute of Astronomy and Astrophysics.  The ARO is partially
supported through the NSF University Radio Observatories program.  The
JCMT was operated by the Joint Astronomy Centre on behalf of the
Science and Technology Facilities Council of the UK, the Netherlands
Organisation for Scientific Research, and the National Research
Council of Canada.  Funding for ongoing CARMA development and
operations was supported by the NSF and CARMA partner universities.
A.~E.~B.\ receives financial support from the Perimeter Institute for
Theoretical Physics and the Natural Sciences and Engineering Research
Council of Canada through a Discovery Grant.  Research at Perimeter
Institute is supported by the Government of Canada through Industry
Canada and by the Province of Ontario through the Ministry of Research
and Innovation.  M.~H.\ acknowledges support from MEXT/JSPS KAKENHI.
L.~L.\ and G.~N.~O.-L.\ acknowledge the financial support of DGAPA,
UNAM, and CONACyT, Mexico.  The EHT gratefully acknowledges equipment
donations from Xilinx Inc.

{\it Facilities:} \facility{EHT}

\appendix

\section{Potential Sources of Error in Closure Phase Estimates}

There are many places where potential errors may be introduced into
VLBI data, including real-world imperfections in the signal chain and
inaccuracies in the input model for correlation.  Many potential
sources of error close.  Among nonclosing errors, some introduce
additional random (zero-mean) error into each closure phase estimate,
while others may introduce biases.  In order to characterize the
significance of our results, we examine which errors might potentially
bias closure phase measurements.

In this appendix we consider potential sources of error from both
theoretical and empirical perspectives.  For the latter, we appeal to
the data themselves to characterize potential biases.  In addition to
the Sgr~A* data reported in this manuscript, two other sources from
the 2013 observations provide high-S/N data to test whether potential
sources of error introduce measurable biases into closure phase
measurements.  The source with the largest correlated flux density on
long baselines in 2013 was BL~Lac, for which a long series of
consecutive scans on day 086 provides a large sample with high
bispectral S/N on all triangles, often exceeding 100 on the FPS and
GJT triangles.  Another bright source that was observed over five
nights in 2013, 3C~279, shows evidence of high polarization and
complicated polarimetric structure on long baselines.  We use data
from these sources to estimate an upper limit of the magnitude of
potential biases by considering matched pairs of simultaneously
measured closure phases from two different data subsets (RCP vs.\ LCP,
high band vs.\ low band, etc.) on both nontrivial and trivial
triangles.

\subsection{Clock and Position Errors}

Each telescope uses a hydrogen maser as its timing and frequency
standard.  The maser signal is very stable over timescales of minutes
but usually exhibits a slow drift over longer timescales.  The
difference between a one pulse-per-second (PPS) signal from the maser
and a PPS signal from the Global Positioning System (GPS) is logged
over time, from which a time offset and drift rate are calculated.
These parameters serve as inputs to the model used for correlation.

The correlator model also takes as inputs the locations of the
telescopes and the celestial coordinates of the source.  The location
of each telescope, including the phase center of the phased arrays,
has been measured with GPS.  Errors on the order of centimeters to
tens of centimeters may be possible due to a combination of GPS
measurement errors and continental drift.  Source coordinates are
obtained from longer-wavelength VLBI catalogues and automatically
corrected for precession.  However, these coordinates may contain
errors on the order of milliarcseconds, and in any case the centroid
of emission at 1.3~mm may be different from that measured at a longer
wavelength due both to frequency-dependent source structure and to
intrinsic source variability.

Because of these effects, as well as a rapidly varying atmosphere, the
a priori correlator parameters are close, though not perfectly
correct, for modelling the delay and rate of a fringe.  Fringe finding
is required post-correlation in order to find residual delays and
rates to compensate for errors in the model.  However, the total
quantities (delay and rate), the sum of the model and residual
quantities, are independent of the input model provided that the a
priori model was within the effective correlator search windows (set
by the accumulation period and spectral resolution).  Thus, small
station-based clock and position errors would not be expected to
introduce biases into the closure phases calculated later in
postprocessing.

\subsection{Polarization Leakage}

The receivers on the EHT telescopes were set up to receive left and
right circularly polarized emission from the sky.  Since no feed or
polarizer (such as a quarter-wave plate) is perfect, a system set up
to receive LCP will nevertheless detect a small portion of RCP
emission, and vice versa.

The full equations for the correlated quantities, including
polarization leakage, are
\begin{eqnarray}
R_1R_2^*  =  G_{1R}G_{2R}^* & [ &
                        (I_{12}+V_{12}) e^{i(-\varphi_1+\varphi_2)} \nonumber \\
  & & + D_{1R} D_{2R}^* (I_{12}-V_{12}) e^{i(+\varphi_1-\varphi_2)} \nonumber \\
  & & + D_{1R}           P_{21}^*       e^{i(+\varphi_1+\varphi_2)} \nonumber \\
  & & + D_{2R}^*         P_{12}         e^{i(-\varphi_1-\varphi_2)}] \and \nonumber \\
L_1L_2^*  =  G_{1L}G_{2L}^*  & [ &
                        (I_{12}-V_{12}) e^{i(+\varphi_1-\varphi_2)} \nonumber \\
  & & + D_{1L} D_{2L}^* (I_{12}+V_{12}) e^{i(-\varphi_1+\varphi_2)} \nonumber \\
  & & + D_{1L}           P_{12}         e^{i(-\varphi_1-\varphi_2)} \nonumber \\
  & & + D_{2L}^*         P_{21}^*       e^{i(+\varphi_1+\varphi_2)}],
\label{pol-equations}
\end{eqnarray}
where numeric subscripts indicate antennas, letter subscripts indicate
the sense of the circularly polarized feed, asterisks indicate complex
conjugation, $G$ indicates complex gain terms, $D$ indicates
polarization leakage terms, $I$ and $V$ indicate Stokes parameters
representing the source total intensity and circular polarization, $P
= Q + iU$ indicates the combination of Stokes parameters representing
the source linear polarization, and $\varphi$ indicates the field
rotation angle \citep{roberts1994}.  The field rotation angle depends
on the mount of the telescope and the location of the receiver; for
some EHT telescopes it is equal to the parallactic angle, and for
others it is the parallactic angle plus or minus the elevation angle.

Sgr~A* exhibits very little ($\sim 1\%$) circular polarization at
1.3~mm \citep{munoz2012,johnson2015b}, so circular polarization would
not be expected to introduce an error of more than a few tenths of a
degree in the closure phase from the terms without leakage ($D$) in
equations~(\ref{pol-equations}).  As the field rotation angles rotate,
the Stokes~I terms vary as the \emph{difference} of the field rotation
angle between the two stations.  In the absence of polarization
leakage, the rotations introduced by the field rotation angles on
three closing baselines cancel, producing no net change in the
measured closure phase.

In contrast, Sgr~A* exhibits high linear polarization at 1.3~mm.  The
median effective linear polarization fraction ($P/I$) ranges from
about 5\% on intrasite and SMT-CARMA baselines to 35\% on the
JCMT-CARMA baseline, with substantial additional variability
\citep{johnson2015b}.  Instrumental polarization leakage ($D$) terms
for the EHT range from 1\% at the SMA to 11\% at the SMT, with a
median value of about 5\%.  Polarization leakage can contaminate the
RR and LL visibilities via the product $DP$, which for Sgr~A* amounts
to a few percent.  This is not large enough for our measured median
closure phase of $6.3\degr$ to be attributable to polarization
leakage.  Leakages vary as the \emph{sum} of the field rotation angles
of the stations on each baseline, with effects on RCP and LCP closure
phases that are approximately symmetric but opposite in sign.

Assuming the median baseline-dependent polarization fractions and
average polarization angles measured for Sgr~A* in 2013 as well as the
$D$-terms derived during instrumental polarization calibration, the
typical biases expected to be introduced by polarization leakage are
$\lesssim 1\degr$ over most of an observing track, with a maximum
effect of about $2\degr$.  The closure phase data on the most
sensitive nontrivial LCP and RCP triangles (FPS and GJT, respectively)
do not have sufficient S/N to be able to detect this difference.  The
trivial DFS and EGT closure phases are consistent with each other to
less than $1\degr$, and the slope of their difference with time is
consistent with zero.  Neither BL~Lac nor 3C~279 shows a statistically
significant bias between the LCP and RCP closure phases on any
triangle (including the nontrivial California-Hawaii-Arizona triangle)
or a slope with time.

\subsection{Gain Errors}

As can be seen in equation~(\ref{pol-equations}), the complex gain
($G$) terms apply to the source term of each visibility as well as all
of the leakage ($D$) terms.  The closure phase is the argument of the
product of three visibilities.  Denoting gain-corrected source
quantities (bracketed terms in equation~(\ref{pol-equations})) by
$[\ldots]$, the measured closure phase is
\begin{eqnarray}
\phi_{123,R} &=& \arg\left((R_1R_2^*)(R_2R_3^*)(R_3R_1^*)\right) \nonumber \\
          &=&  \arg\left((G_{1R}G_{2R}^*)(G_{2R}G_{3R}^*)(G_{3R}G_{1R}^*)
               [R_1R_2^*][R_2R_3^*][R_3R_1^*]\right) \nonumber \\
          &=&  \arg\left([R_1R_2^*][R_2R_3^*][R_3R_1^*]\right).
\end{eqnarray}
Thus, station-based, frequency-independent complex gain errors do not
introduce additional closure phase errors beyond those already present
due to polarization leakage.

\subsection{Bandpass Effects}

Frequency-dependent bandpass errors can introduce closure phase
errors.  The bandpass response of a telescope is in general a complex
quantity, containing both amplitude and phase structure.

HOPS provides partial mitigation of bandpass effects.  The amplitude
of each station is automatically normalized to the autocorrelation
response on a per-channel level.  This can theoretically still result
in small errors when the autocorrelation response does not match the
desired crosscorrelation response---for instance, if the IF response
reduces sensitivity to sky signals near the edge of the band, or if
subchannel structure in the bandpass response is significant.

EHT telescopes do not have a pulse calibration system to align the
phases of each channel.  Instead, the EHT data reduction uses manual
instrumental phase calibration on very bright sources to derive phase
offsets to apply to each channel on a per-station basis.  The signal
for all frequencies within each 480~MHz band passes through the same
chain of electronics before being sampled by a single backend.  As a
result, the manual phases required to flatten the phase structure of
the bandpass vary smoothly across the band.  Before they are removed,
channel-to-channel phase differences are typically a few degrees to
$20\degr$, with larger differences sometimes seen at the edges of the
band.

Data from BL~Lac indicate that any biases introduced by bandpass
effects do not exceed a few tenths of a degree.  High-band and
low-band closure phases on nontrivial triangles agree to less than a
degree.  An additional test in which the low band was subdivided into
two pieces (the first 8 and the last 7 channels) found consistent
closure phases on all triangles at the level of $0.2\degr$ or better.

\end{document}